\documentstyle[12pt]{article}
\newcommand{\BE}{\begin{equation}}
\newcommand{\EE}{\end{equation}}
\newcommand{\BD}{\begin{displaymath}}
\newcommand{\ED}{\end{displaymath}}

\newcommand{\ed}{{\rm d}}
\newcommand{\dc}[2]{\left(\begin{array}{c} #1 \\ #2 \end{array}\right)}

\newcommand{\prl}[3]{{\it Phys. Rev. Lett.} {\bf #1}, #2 (#3)}

\newcommand{\cmp}[3]{{\it Comm. of Math. Phys.} {\bf #1}, #2 (#3)}

\newcommand{\apny}[3]{{\it Ann. of Phys.} (N.Y.) {\bf #1}, #2 (#3)}
\newcommand{\ap}[3]{{\it Ann. of Phys.} {\bf #1}, #2 (#3)}
\newcommand{\ijmpa}[3]{{\it Inter. J. of Mod. Phys.} A{\bf #1}, #2 (#3)}
\newcommand{\plb}[3]{{\it Phys. Lett.} B{\bf #1}, #2 (#3)}

\newcommand{\book}[3]{#1,`{\it #2}' (#3)}

\title{\bf Canonical Chern-Simons Theory and the Braid Group on a Riemann
Surface
\thanks{Submitted to Physics Letters B}
}

\author{Mario Bergeron \thanks{This work is supported in part by a UBC
Fellowship and FCAR.} , David Eliezer and
Gordon Semenoff \thanks{This work is supported in part by the Natural Sciences
and Engineering Research Council of Canada.}
\\ {\it Department of Physics}\\ {\it University of British Columbia}
\\ {\it Vancouver, British Columbia, Canada  V6T 1Z1}}

\begin{document}

\maketitle

\begin{abstract}
We find an explicit solution of the Schr\"odinger equation for a
Chern-Simons theory coupled to charged particles on a Riemann surface,
when the coefficient of the Chern-Simons term is a rational number
(rather than an integer) and where the total charge is zero.  We find
that the wave functions carry a projective representation of the group
of large gauge transformations. We also examine the behavior of the
wave function under braiding operations which interchange particle
positions.  We find that the representation of both the braiding
operations and large gauge transformations involve unitary matrices
which mix the components of the wave function.  The set of wave
functions are expressed in terms of appropriate Jacobi theta functions.
The matrices form a
finite dimensional representation of a particular (well known to
mathematicians) version of the braid group on the Riemann surface.  We
find a constraint which relates the charges of the particles, $q$, the
coefficient of the Chern-Simons term, $k$ and the genus of the
manifold, $g$: $q^2(g-1)/k$ must be an integer.  We discuss a duality
between large gauge transformations and braiding operations.
\end{abstract}

\newpage

It is by now well established that particles confined to a two
dimensional space can have fractional statistics.  Such particles are
called anyons.  Interest in them is partially motivated by their
physical effects such as their conjectured role in the fractionally
quantized Hall effect or high temperature superconductivity, and
partially by the fact that the description of anyons uses interesting
mathematical structures.  Anyons are a generalization of ordinary
bosons or fermions where the wave functions of many identical
particles, instead of being symmetric or antisymmetric, carry a
representation of the braid group on the two dimensional space.  This
is often described mathematically by coupling the currents of the
particles to the gauge field of a Chern-Simons theory,
\BE \label{csa}
S=-{k \over 4 \pi}\int A\ed A+\int A_\mu j^\mu \ed^3 x
\EE
The solutions of this field theory take the form of a set of wave
functions which at a given time depend on the particle positions.  It
is well known that these wave functions carry a non-trivial
representation of the braid group, where the representation is
specified by the parameter $k$ as well as the number of particles and
the topology of the 2 dimensional space.  The braid group is an
infinite discrete non-Abelian group and has many potentially
interesting representations (see, for example \cite{4}).

Abelian Chern-Simons theory coupled to classical point particles was
solved on the plane by Dunne, Jackiw and Trugenberger \cite{8}.  In
that case there are no degrees of freedom, the Hilbert space is
one-dimensional and the only quantum state is given by a single
unimodular complex number. The phase of the complex number depends on
the parameters of the model, i.e. the coupling constant $k$ and the
histories of time evolution of the positions of the particles in such
a way that if we consider a periodic trajectory, $z_i(t),~
i=1,\ldots,N$, $t\in[0,1]$ and $z_i(1)=z_i(0)$ the phase of the wave
function changes by the well-known factor $$
\frac{1}{ k}\sum_{i<j}\int_0^1 dt\frac{d}{dt}
{\rm Im}\ln\left( z_i(t)-z_j(t) \right) $$ which counts the changes of
relative angles of positions of the particles.  This can be
interpreted as the wave function carrying a one--dimensional unitary
representation of the Nth order braid group on the plane.  The
statistics parameter is the change in phase of the wave function under
interchange of two particle positions.  Here it is given by $\pi\over
k$.

In another interesting paper, Bos and Nair \cite{3} have solved the
Schr\"odinger equation for Abelian Chern-Simons theory coupled to
particles when the space is a Riemann surface of genus $g$ and when
$k$, the coefficient of the Chern-Simons term, is an integer.  In this
case the representation of the braid group which is obtained is more
interesting.  It is generally multi-dimensional and represents the
braid group on the Riemann surface which has more (non--Abelian)
structure relations than the braid group on the plane.  In this Letter
we shall show how this solution generalizes to Chern-Simons theory
with $k$ a rational number and examine the resulting properties of the
braid group representations which we obtain, realizing explicitly the
representations found in \cite{4}.

In Abelian Chern-Simons theory, unlike the non-Abelian case, there is
no reason why $k$ should be quantized but could in principle be any
real number.  Polychronakos \cite{6} has shown that, even on a torus,
$k$ need not be quantized.  Furthermore, Witten \cite{2} has
argued that the dimension of the Hilbert space of pure Chern-Simons
theory on a Riemann surface with genus $g$ is $k^g$ when $k$ is an
integer.  Polykronakos \cite{6} showed that when $k$ is rational, the
dimension of the Hilbert space of pure Chern-Simons theory is
$(k_1k_2)^g$.  It is also known that, when $k$ is an integer, the
representations of the braid group on a Riemann surface obtained by
coupling to point particles depends on $k$ and $g$ \cite{4}.

We start from Abelian Chern-Simons theory coupled to the current due
to a gas of charge particles (\ref{csa}).  We shall first demonstrate
how to extract the physical degrees of freedom from the non-trivial
cohomology classes of the gauge field and the complex structure of the
Riemann surface. Later, invariance under large gauge transformations
and modular transformations will be used to show that the Hilbert
space is finite dimensional.

Before quantizing, we rewrite (\ref{csa}) in term of topological
quantities on the 2-dimensional Riemann surface, ${\cal M}$, of genus
g.  We decompose $A$ into its exact, coexact and harmonic parts. More
precisely, the Hodge decomposition of $A$, on ${\cal M}$, is given by
($\ed$ and ${}^*$ act on ${\cal M}$ in this paper)
\BE \label{hda}
A=\ed(\frac{1}{\Box'}{}^*\ed^*A)+{}^*\ed(\frac{1}{\Box'}{}^*\ed
A)+\frac{2\pi i}{k}\sum_{l=1}^g(\bar\gamma_l \omega^l-\gamma_l\bar\omega^l)
\EE
where $1/\Box'$ is the inverse of the laplacian $(\Box)$ acting on
0-forms where the prime means that the zero modes are removed. The
zero modes of $\ed$ and ${}^*\ed$ (or $\Box$) acting on a one-form are
spanned by the set of Abelian differentials, $\omega^l$, on ${\cal
M}$.

We can represent the homology of ${\cal M}$ in terms of generators
$a_l$ and its conjugate generators $b_l$, $l=1,\ldots,g$.
The intersection numbers of these generators are given by
\BE \label{inter}
\nu(a_l,a_m)=\nu(b^l,b^m)=0,\ \nu(a_l,b^m)=-\nu(b^m,a_l)=\delta_l^m
\EE
where $\nu(C_1,C_2)$ is the signed intersection number (number of
right-handed minus number of left handed crossings) of the oriented
curves $C_1$ and $C_2$.  The holomorphic harmonic one-forms $\omega^i$
have the standard normalization
\cite{1}
\BD
\oint_{a_l}\omega^m=\delta_l^m,\ \oint_{b^l}\omega^m=\Omega^{lm}
\ED
The matrix $\Omega^{lm}$ is symmetric and its imaginary part is
positive definite. This actually defines a metric in the space of
holomorphic harmonic forms
\BE \label{hfn}
 i\int_{\cal M}\omega^l\wedge\bar\omega^m=2{\rm
Im}(\Omega^{lm})=G^{lm},\ G_{lm}G^{mn}=\delta_l^n
\EE
We will use $G_{lm}$ and $G^{lm}$ to lower or raise indices when needed and use
Einstein summation convention over repeated indices.

Any linear relation, with integer coefficients, of $a_l$ and $b_l$
that satisfy (\ref{inter}) is another valid basis for the homology
generators.  These transformations form a symmetry of the Chern-Simons
theory and comprise the modular group, $Sp(2g,Z)$:
\BE \label{mt}
\dc{a}{b}\rightarrow S\dc{a}{b}\qquad{\rm where}\qquad S=\dc{D\ C}{B\ A}
\EE
with $SES^\top=E$ and $E=\dc{\ 0\ 1}{-1\ 0}$.  The $g\times g$
matrices $A,B,C,D$ have integer entries.

We can define
\BD
\xi=-\frac{k}{2\pi}\frac{1}{\Box'}{}^*\ed^* A~,~~\ F={}^*\ed A
\ED
So we then have (including the $A_0$ term)
\BE \label{ta}
A=A_0\ed t-\frac{2\pi}{k}\ed\xi+{}^*\ed(\frac{1}{\Box'}F)+2\pi i
(\bar\gamma_l \omega^l-\gamma_l\bar\omega^l)
\EE
Similarly we can write the current ${\bf
j}=j^\mu\frac{\partial}{\partial x^\mu}$ into a one-form $j_\mu\ed
x^\mu=j_0\ed t+\tilde j$, using the metric which we assume to be flat
in the time direction ($g_{00}=1,\ g_{01}=g_{02} =0$ and the remaining
components forming the metric on ${\cal M}$). We can use again the
Hodge decomposition of ${}^*\tilde j$ on ${\cal M}$
\BE \label{tj}
{}^*\tilde j=-\ed\chi+{}^*\ed\psi+i(j_l\bar\omega^l-\bar j_l\omega^l)
\EE
The continuity equation $\frac{\partial j_0}{\partial t}\ed^3x+\ed^*\tilde j
\wedge\ed t=0$ can
be used to solve for $\psi$
\BD
\psi=-\frac{1}{\Box'}\frac{\partial j_0}{\partial t}
\ED

We shall consider a set of point charges moving on ${\cal M}$, with
trajectories $z_i(t)$ and charge $q_i$, where $z_i(t)\ne z_j(t)$ for
$i\ne j$. For technical reasons (involving the solution of Gauss' law)
we shall assume here that the total charge is zero, $\sum_i q_i=0$.
We believe that this is only a technical restriction and can be
generalized.  We shall discuss this generalization, the inclusion of the
nonzero chrge sector, in a future
publication.\footnote{Gauss' Law (see ahead (\ref{glc})) implies
that the (single) additional degree of freedom associated with the
nonzero charge should commute with those in
the zero charge sector.  This (and the additive structure of the
Hamiltonian) would imply that the full wave function
is a product of the zero charge wave function (see ahead equation
(\ref{wvfn})) and an additional factor for the nonzero charge,
depending only on the single quantum variable $\xi_0$ conjugate to $\int F$,
and fixed by Gauss' Law to be $\exp{-{2 \pi i \over k}  Q \xi_0}$.
Thus our results will be largely anaffected by this inclusion of
the zero charge sector.}  The current is represented by
\BE \label{pcc}
j_0(z,t)=\sum_i q_i\delta(z-z_i(t)),\ \tilde j(z,t)=\sum_i
q_i\delta(z-z_i(t))
\frac{1}{2}(\dot z_i(t)\ed\bar z+\dot{\bar z}_i(t)\ed z)
\EE
Integrating (\ref{pcc}) with the harmonic forms $\omega^l$ , we find
the topological components of the current in (\ref{tj})
\BE \label{tc}
j^l(t)=\sum_i q_i\dot z_i(t)\omega^l(z_i(t))\ ,\qquad\bar j^l(t)=\sum_i q_i
\dot{\bar z}_i(t)\bar \omega^l(\bar z_i(t))
\EE

To solve for $\chi$, it is best to use complex notation
\BD
R=\psi+i\chi=R(z,\bar z)
\ED
where we find ${}^*\ed\chi+\ed\psi=\partial_z\bar R\ed z+\partial_{\bar z}
R\ed\bar z$. From (\ref{tj}), (\ref{pcc}) and using (\ref{tc}) we find
\BD
\partial_z\bar R+\bar j_l\omega^l(z)=\frac{1}{2}\sum_i q_i\dot{\bar z}_i
\delta(z-z_i(t))
\ED
\BE \label{R}
\partial_{\bar z} R+j_l\bar\omega^l(\bar z)=\frac{1}{2}\sum_i q_i\dot{z}_i
\delta(z-z_i(t))
\EE
To solve (\ref{R}) for $R$, we will need the prime form
\BD
E(z,w)=(h(z)h(w))^{-{\frac{1}{2}}}\cdot\Theta\dc{1/2}{1/2}(\int_z^w\omega|
\Omega)
\ED
where $h(z)=\frac{\partial}{\partial u^l}\Theta\dc{1/2}{1/2}
(u|\Omega)|_{u=0}\cdot\omega^l(z)$. The prime form is antisymetric in
the variables $z$ and $w$ and behaves like $z-w$ when $z\approx w$
(the $h(z)$ which appear in the denominator are for
normalization).\footnote{This formalism can also be extended to
include the sphere, where there are no harmonic 1-forms at all (the
space of cohomology generators has dimension zero) by properly
defining the prime form. We use stereographic projection to map the
sphere into the complex plane and use $E(z,w)=z-w$ as the definition
of the prime form. }

The theta functions \cite{7} are defined by
\BE \label{thetaf}
\Theta\dc{\alpha}{\beta}(z|\Omega)=\sum_{n_l}e^{i\pi(n_l+\alpha_l)\Omega^{lm}
(n_m+\alpha_m)+2\pi i(n_l+\alpha_l)(z^l+\beta^l)}
\EE
where $\alpha,\ \beta\in [0,1]$, have the following property
\BD
\Theta\dc{\alpha}{\beta}(z^m+
s^m+\Omega^{ml}t_l|\Omega)=e^{2\pi i\alpha_l s^l- i\pi
t_m\Omega^{ml}t_l-2\pi i t_m(z^m+\beta^m)}\Theta\dc{\alpha}{\beta}(z|
\Omega)
\ED
for integer--valued vectors $s^m$ and $t_l$.  For a non-integer
constant $c$
\BD
\Theta\dc{\alpha}{\beta}(z^m+c
\Omega^{ml}t_l|\Omega)=e^{-i\pi c^2 t_m\Omega^{ml}
t_l-2\pi i c t_m(z^m+\beta^m)}\Theta\dc{\alpha-ct}{\beta}(z|\Omega)
\ED

The solution of (\ref{R}) is
\BE
R=\frac{\partial}{\partial t}[-\frac{1}{2\pi}\sum_i
q_i\log(\frac{E(z,z_i(t))}
{E(z_0,z_i(t))})]-j_l(t)\int_{z_0}^z(\bar\omega^l-\omega^l)
\EE
where we have chosen $R$ such that $R(z_0,\bar z_0)=0$ for an
arbitrary point $z_0$ (We can choose $z_0=\infty$ for genus zero).
The important fact about $R$ is that it is a single-valued function.
If we move $z$ around any of the homology cycles, $R$ returns to its
original value. In fact, this is also true for windings of $z_0$, an important
relation since it is only a reference point.  So
\BE \label{chi}
\chi=\frac{\partial}{\partial t}[-\frac{1}{2\pi}\sum_i q_i
{\rm Im}\log(\frac{E(z,z_i(t))}{E(z_0,z_i(t))})]+
\frac{i}{2}[(j_l(t)+\bar j_l(t))\int_{z_0}^z(\bar\omega^l-\omega^l)]
\EE

Now we are ready to solve the Chern-Simons theory on ${\cal M}$.  By
putting (\ref{ta}) and (\ref{tj}) back into (\ref{csa}) we find
\BD
S=\frac{1}{2}\int(\xi P\dot F-\dot\xi PF)\ed^3x+i\pi k\int(\gamma^l
\dot{\bar\gamma}_l-\dot\gamma^l\bar\gamma_l)\ed t+\int A_0(j_0
-\frac{k}{2\pi}F)\ed^3x
\ED
\BE \label{tcsa}
-\int(\frac{2\pi}{k}\xi\frac{\partial j_0}{\partial t}+FP\chi)\ed^3x
+2\pi i\int(j_l\bar\gamma^l -\bar j^l\gamma_l)\ed t
\EE
where $P$ is a projection operator onto the space orthogonal to the
zero mode.  From this we obtain the equal-time commutation relations
of the quantum theory
\BE \label{qv}
[\xi(z),F(w)]=-iP\delta(z-w)\qquad {\rm or}\qquad
F(z)=iP\frac{\delta}{\delta
\xi(z)}
\EE
and
\BE \label{tqv}
[\gamma_l,\bar\gamma_m]=-\frac{1}{2\pi k}G_{lm}\qquad {\rm or}\qquad
\bar\gamma_l=\frac{1}{2\pi k}G_{lm}\frac{\partial}{\partial\gamma_m}=
\frac{1}{2\pi k}\frac{\partial}{\partial\gamma^l}
\EE
The projection operator in (\ref{qv}) changes the delta function to
$\delta(z-w)-1/{\rm Area} ({\cal M})$. The functional derivative must
also be defined using this projection operator.  With this holomorphic
polarization \cite{3} it is convenient to use the following measure in
$\gamma$ space
\BD
(\Psi_1|\Psi_2)=\int e^{-2\pi k\gamma^m G_{ml}\bar\gamma^l}
\Psi_1^*(\bar\gamma)\Psi_2(\gamma)|G|^{-1}\prod_m \ed\gamma^m\ed\bar\gamma^m
\ED
where $|G|=\det(G_{mn})$. With this measure, we find that $\gamma^\dagger=\bar
\gamma$ as it should be.

$A_0$ is a Lagrange multiplier which enforces the Gauss' law
constraint
\BE \label{glc}
F(z)-\frac{2\pi}{k}j_0(z)=iP\frac{\delta}{\delta\xi(z)}-\frac{2\pi}{k}j_0(z)
\approx 0
\EE
Note that since the total charge is zero, the total magnetic flux
should be zero, as is necessary on a compact space when the gauge
field $A$ is a globally defined 1-form.

Under a modular transformation, the basis $\gamma^l$, ${\bar\gamma}^l$
will be transformed accordingly. This will not change the choice of
polarization, since the modular transformations do not mix $\gamma$
and $\bar\gamma$.

{}From (\ref{tcsa}), (\ref{qv}) and (\ref{tqv}), we find that the hamiltonian,
in the $A_0=0$ gauge, can be separated into two commuting parts
\BD
H=-\int_{\cal M}A\wedge^*\tilde j= H_0+H_T
\ED
where
\BE
H_0=\int_{\cal M}(\frac{2\pi}{k}\xi\frac{\partial j_0}{\partial t}+
i\chi P\frac{\delta}{\delta\xi})\ed^2 x
\EE
(note that $\ed^2x=-\frac{i}{2}\ed z\wedge\ed\bar z$) while the
additional part that takes care of the topology is
\BE \label{th}
H_T=i(2\pi\bar j_l\gamma^l-\frac{1}{k}j^l\frac{\partial}
{\partial\gamma^l})
\EE

To solve the Schr\"odinger equation, we will use the fact that the
hamiltonian separates, thus writing the wave function as
\BD
\Psi(\xi,\gamma,t)=\Psi_0(\xi,t)\Psi_T(\gamma,t)
\ED
with the Gauss law constraint (\ref{glc})
\BD
(iP\frac{\delta}{\delta\xi}-\frac{2\pi}{k}j_0)\Psi_0(\xi,t)=0
\ED
which is solved by
\BE \label{gwf}
\Psi_0(\xi,t)=\exp[-\frac{2\pi i}{k}\int_{\cal M}\xi(z) j_0(z,t)\ed^2 x]
\Psi_c(t)
\EE

The first Schr\"odinger equation is
\BD
i\frac{\partial\Psi_0(\xi,t)}{\partial t}=H_0 \Psi_0(\xi,t)=
\left[ \int_{\cal M} \left( \frac{2\pi}{k}\xi\frac{\partial j_0}{\partial t}+
i\chi P\frac{\delta}{\delta\xi} \right) \ed^2 x\right] \Psi_0(\xi,t)
\ED
which has the solution \cite{2}
\BE \label{psij}
\Psi_c(t)=\exp\left[ -\frac{2\pi i}{k} \int_0^t \int_{\cal M}
\chi(z,t^\prime) j_0(z,t^\prime) \ed^2 x \ed t^\prime \right]
\EE
For a system of point charges, the use of (\ref{chi}) allows us to write
(\ref{gwf}) and (\ref{psij}) as
\BE \label{bgp}
\Psi_0(\xi,t)=\exp\left[-\frac{2\pi i}{k}\sum_i q_i\xi(z_i(t))+\frac{i}{2k}
\sum_{ij} q_i q_j \int_0^t \ed t\dot\theta_{ij}(t)+\Phi(t) \right]
\EE
where
\BD
\Phi(t)=\frac{\pi}{k}\left[\int_0^t j_l(t^\prime)\ed t^\prime \int_0^
{t^\prime} \bar j^l(t^{\prime\prime})\ed t^{\prime\prime}-\int_0^t\bar
j_l( t^\prime)\ed t^\prime \int_0^{t^\prime} j^l(t^{\prime\prime})\ed
t^{\prime
\prime}\right]
\ED
\BE \label{phase}
+\frac{\pi}{2k}\left[\int_0^t \bar j_l(t^\prime)\ed t^\prime \int_0^ t
\bar j^l(t^\prime)\ed t^\prime-\int_0^t j_l( t^\prime)\ed t^\prime
\int_0^t j^l(t^\prime)\ed t^\prime\right]
\EE
and
\BD
\theta_{ij}(t)={\rm Im}\log\left[\frac{E(z_i(t),z_j(t))}{E(z_i(t),z_0)E(z_0
,z_j(t))}\right]
\ED
\BE \label{angfunc}
+{\rm Im}\left[\int_{z_0}^{z_i(0)}\omega^l\int_{z_j(0)}^{z_j(t)}
(\omega_l+\bar\omega_l)+\int_{z_0}^{z_j(0)}\omega^l\int_{z_i(0)}^{z_i(t)}
(\omega_l+\bar\omega_l)\right]
\EE
is a multi-valued function defined using the prime form.  We will need
the phase (\ref{phase}) for the topological part of the wave function.
The function $\theta_{ij}(t)$ is the angle function for particle $i$
and $j$. For $i=j$, we choose a framing $z_i(t)=z_j(t)+\epsilon
f_i(t)$ which lead to the replacement of $E(z_i(t),z_i(t))$ by
$f_i(t)$.
So the wave function (\ref{bgp}), with the angle function (\ref{angfunc}),
accurately forms an Abelian representation of the braid group \cite{2,8}.

Now, the topological part of the hamiltonian is used to find the part of the
wave function affected by the currents going around the non-trivial loops of
${\cal M}$. The Schr\"odinger equation for (\ref{th}) is
\BD
i\frac{\partial\Psi_T(\gamma,t)}{\partial t}=
H_T\Psi_T(\gamma,t)=
i\left(2\pi\bar j_l\gamma^l -\frac{1}{k} j^l
\frac{\partial}{\partial\gamma^l} \right)
\Psi_T(\gamma,t)
\ED
which has the solution
\BE
\Psi_T(\gamma,t)
=\exp \left[ 2\pi\gamma^l \int_0^t\bar j_l(t^\prime)
\ed t^\prime - \frac{2\pi}{k} \int_0^t j_l(t^\prime)
\ed t^\prime \int_0^{t^\prime} \bar j^l(t^{\prime\prime})\ed t^{\prime\prime}
\right] \tilde\Psi_T(\gamma,t)
\EE
Note that with the phase (\ref{phase}), the double integral above will turn
into $\int_0^t j_l(t^\prime)\ed t^\prime \cdot\int_0^t \bar j^l(t^{\prime})
\ed t^{\prime}$, a topological expression.

The remaining equation for $\tilde\Psi_T(\gamma,t)$
\BE
\frac{\partial\tilde\Psi_T(\gamma,t)}{\partial t}
=-\frac{1}{k}\ j^l \frac{\partial\tilde\Psi(\gamma,t)}
{\partial\gamma^l}
\EE
is easily solved in the form
\BE \label{tpsit}
\tilde\Psi_T(\gamma^l,t)
=\tilde\Psi_T(\gamma^l -\frac{1}{k} \int_0^t j^l(t^\prime)\ed t^\prime)
\EE

The wave function (\ref{tpsit}) is not arbitrary, but must satisfy the
invariance of the action (\ref{csa}) under large gauge transformations, when
there is no current around. So let us set $j^\mu=0$ for a while and find the
condition on $\tilde\Psi_T$.

In general, the large $U(1)$ gauge transformations are given by the set of
single-valued gauge functions, with $s^m$ and $t_m$ integer-valued vectors,
\BD
U_{s,t}(z)=\exp\left(2\pi i(t_m\eta^m(z)-s^m\tilde\eta_m(z)\right)
\ED
where
\BD
\eta^m(z)=i\int_{z_0}^z(\bar\Omega^{ml}\omega_l-\Omega^{ml}\bar\omega_l)\
,\qquad \tilde\eta_m(z)=-i\int_{z_0}^z(\omega_m-\bar\omega_m)
\ED

If we change the endpoint of integration by $z\rightarrow z+a_l u^l+b^m v_m$
with $u,v$ integer and $a,b$ defined in
(\ref{inter}), we find $\eta^m\rightarrow\eta^m+u^m,\
\tilde\eta_m\rightarrow\tilde\eta_m+v_m$ and $U_{s,t}\rightarrow U_{s,t}e^
{2\pi i(t_m u^m-
s^m v_m)}=U_{s,t}$. The transformation of the gauge field (\ref{hda}) under
$U_{s,t}$ is given by
\BE \label{lgt}
\gamma^m\rightarrow\gamma^m+s^m+\Omega^{ml}t_l\ ,\qquad\bar\gamma^m\rightarrow
\bar\gamma^m+s^m+\bar\Omega^{ml}t_l
\EE
The classical operator that produces the transformation (\ref{lgt})
\BD
c_{s,t}(\gamma,\bar\gamma)=\exp\left[(s^m+\Omega^{ml}t_l)\frac{\partial}
{\partial\gamma^{m}}+(s^m+\bar\Omega^{ml}t_l)\frac{\partial}{\partial\bar
\gamma^{m}}\right]
\ED
must be transformed into the proper quantum operator acting on the wave
function
$\tilde\Psi_T$. By using the commutation (\ref{tqv}) to replace $\frac
{\partial}{\partial\bar\gamma^{m}}$ by $-2\pi k\gamma_m$ we find the
operators $C_{s,t}$ which implement the large gauge transformations \cite{5}
\BE \label{glgt}
C_{s,t}(\gamma)=\exp\left[-2\pi k(s^m+\bar\Omega^{ml}t_l)\gamma_m-\pi
k(s^m+\bar
\Omega^{ml}t_l)G_{mn}(s^n+\Omega^{nl}t_l)\right]e^{(s^m+\Omega^{ml}t_l)\frac
{\partial}{\partial\gamma^{m}}}
\EE

The quantum operators $C_{s,t}$ do not commute among themselves for
non-integer $k$. From now on we will set $k=\frac{k_1}{k_2}$ for integer
$k_1$ and $k_2$. Now, in contrast with their classical counterparts, the
operators $C_{s,t}$ satisfy the clock algebra
\BE \label{clka}
C_{s_1,t_1}C_{s_2,t_2}=e^{-2\pi ik(s^m_1{t_m}_2-s^m_2{t_m}_1)}C_{s_2,t_2}
C_{s_1,t_1}
\EE
Their action on the wave function is
\BD
C_{s,t}(\gamma)\tilde\Psi_T(\gamma^m)=\exp\left[-2\pi k(s^m+\bar\Omega^{ml}t_l)
\gamma_m\right.
\ED
\BE \label{ca}
\left.-\pi k(s^m+\bar\Omega^{ml}t_l)G_{mn}(s^n+\Omega^{nl}t_l)\right]
\tilde\Psi_T(\gamma^m+s^m+\Omega^{ml}t_l)
\EE
On the other hand $C_{k_2s,k_2t}$ commutes with everything and must be
represented only by phases $\phi_{s,t}$. This implies, using (\ref{ca}),
\BD
\tilde\Psi_T(\gamma^m+k_2(s^m+\Omega^{ml}t_l))=\exp\left[-i\phi_{s,t}+2\pi
k_1(s^m+\bar\Omega^{ml}t_l)\gamma_m\right.
\ED
\BE \label{algc}
\left.+\pi k_1k_2(s^m+\bar\Omega^{ml}t_l)G_{mn}
(s^n+\Omega^{nl}t_l)\right]\tilde\Psi_T(\gamma^m)
\EE
The only functions that are doubly (semi-)periodic are combinations of the
theta functions (\ref{thetaf}).
After some algebra, we find that the set of functions
\BE \label{tqf}
\Psi_{p,r}\dc{\alpha}{\beta}(\gamma|\Omega)=e^{\pi k\gamma^m\gamma_m}\Theta
\dc{\frac{\alpha+k_1 p+k_2 r}{k_1 k_2}}{\beta}(k_1\gamma|k_1 k_2\Omega)
\EE
where $p=1,2,\dots,k_2$ and $r=1,2,\dots,k_1$ with $\alpha,\ \beta\in[0,1]$
solve the above algebraic conditions (\ref{algc}).
Their inner product is given by
\BE
(\Psi_{p_1,r_1}|\Psi_{p_2,r_2})=\int_P e^{-2\pi k\gamma^mG_{ml}\bar\gamma^l}
\overline{\Psi_{p_1,r_1}(\gamma)}\Psi_{p_2,r_2}(\gamma)
|G|^{-1}\prod_m \ed\gamma^m\ed\bar\gamma^m
\EE
\BD
=|G|^{-\frac{1}{2}}\delta_{p_1,p_2}\delta_{r_1,r_2}
\ED
The integrand is completely invariant under the translation (\ref{lgt}), thus
we restrict the integration to one of the
plaquettes $P$ ($\gamma^m=u^m+\Omega^{ml}v_l$ with $u,v\in[0,1]$), the phase
space
of the $\gamma$'s.

Under a large gauge transformation
\BD
C_{s,t}\Psi_{p,r}\dc{\alpha}{\beta}(\gamma)=e^{2\pi ikp_m s^m+i\pi ks^m t_m+
\frac{2\pi i}{k_2}(\alpha_m s^m-\beta^m t_m)}\Psi_{p+t,r}\dc{\alpha}{\beta}
(\gamma)
\ED
\BE
=\sum_{p^\prime}[C_{s,t}]_{p,p^\prime}\Psi_{p^\prime,r}\dc{\alpha}{\beta}
(\gamma)
\EE
The matrix $[C_{s,t}]_{p,p^\prime}$ forms a $(k_2)^g$ dimensional
representation
of the algebra (\ref{clka}) of large gauge transformations.

The parameters $\alpha$ and $\beta$ appear as free parameters, but in fact
they may be fixed such that we obtain a modular invariant wave function.
The modular transformation (\ref{mt}) on our set of functions (\ref{tqf}) is
\BD
\Psi_{p,r}\dc{\alpha}{\beta}(\gamma|\Omega)\rightarrow|C\Omega+D|^{-\frac{1}
{2}}e^{-i\pi\phi}\Psi_{p,r}\dc{\alpha^\prime}{\beta^\prime}(\gamma
^\prime|\Omega^\prime)
\ED
where $\gamma^\prime={(C\Omega+D)^{-1}}^\top \gamma,\ \Omega^\prime=(A\Omega+B)
(C\Omega+D)^{-1}$ and $\phi$ is a phase that will not concern us here (and
$G^\prime_{lm}=[(C\Omega+D)^{-1}]_{lr}G_{rs}[(C\bar\Omega+D)^{-1}]_{sm}$).
Most important are the new variables
\BD
\alpha^\prime=D\alpha-C\beta-\frac{k_1 k_2}{2}(CD^\top)_d\qquad\beta^\prime=
-B\alpha+A\beta-\frac{k_1 k_2}{2}(AB^\top)_d
\ED
where $(M)_d$ mean $[M]_{dd}$, the diagonal elements.

A set of modular invariant wave functions \cite{4,5,6} can exist only when
$k_1 k_2$ is even, where we set $\alpha=\beta=0$ (and also $\phi=0$).
In the case of odd $k_1 k_2$, we can set $\alpha,\ \beta$ to either $0$ or
$\frac{1}{2}$, which amount to the addition of a spin structure on the wave
functions. This will increase the number of functions by $4^g$ which will now
transform non trivially under modular transformations.

Considering a set of point charges leads to the set of wave functions
\BD
\Psi_{p,r}\dc{\alpha}{\beta}(\xi,\gamma,t|\Omega)=\exp\left[\pi k\gamma^m
\gamma_m+2\pi\gamma^m\int_0^t(\bar j_m-j_m)\ed t^\prime-\frac{2\pi i}{k}\sum_i
q_i\xi(z_i(t))\right.
\ED
\BD
\left.+\frac{i}{2k}\sum_{ij}q_iq_j(\theta_{ij}(t)-\theta_{ij}(0))+
\frac{\pi}{2k}\int_0^t(j_m-\bar j_m)\ed t^\prime\cdot\int_0^t(j^m-
\bar j^m)\ed t^\prime\right]
\ED
\BE \label{wvfn}
\cdot\Theta\dc{\frac{\alpha+k_1 p+k_2 r}{k_1 k_2}}{\beta}(k_1\gamma^m-k_2
\int_0^tj^m\ed t^\prime|k_1 k_2\Omega)
\EE

The wave function depends on charge
positions through the integrals over the topological components of the
current $j^m, \bar j^m$, and through the function $\theta_{ij}(t) -
\theta_{ij}(0)$.  Consider for a moment motions of the particles
which are closed curves, and are homologically trivial.  We focus
first on the integrals over  $j^m, \bar j^m$.  If, for example a single
particle moves in a circle,
we find that the integral of these topological currents vanishes, we
conclude that these currents contribute nothing additional to the phase of
the wave function under these kinds of motions.  The function  $\theta_{ij}
(t) - \theta_{ij}(0)$ must be treated differently here, because it has
singularities when particles coincide, and thus, while motions that
encircle no other particles may be easily integrated to get zero, this
is not true when other particles are enclosed by one of the particle
paths, and the result is nonzero in this case, in fact it is $2\pi$.
Nevertheless, this function is still independent of the particular
shape of the particle path. Actually the definition of $\theta_{ij}$ in term
of the prime form $E(z,w)$ is just the generalization to an arbitrary Riemann
surface of the well known angle function on the plane, that is as the angle of
the line joining the particle $i$ and $j$ compare to a fixed axis of
reference, determined by $z_0$ here.
Thus, we may conclude that, under permutations of particles of charge
$q$, the
wave functions defined here acquire the phase $\sigma = e^{\frac{i\pi}{k}q^2}$.

For homologically nontrivial motions of a single particle on ${\cal M}$
, the current integral  $\int_0^t
j^l(t^\prime)\ed t^\prime$ will in general change as $\int_0^t
j^l(t^\prime)\ed t^\prime \longrightarrow  \int_0^t
j^l(t^\prime)\ed t^\prime +  s^l+\Omega^{lm}t_m$, where $s^l$ and $t_m$
are integer-valued vectors whose entries denote the number of windings
of the particle around each homological cycle.
However, now,
for multi-particle non-braiding paths, there is no contribution comming from
$\theta_{ij}$.
Thus, the wave functions become
\BD
\Psi_{p,r}(t)=\exp\left[-\frac{2\pi i}{k}r_m s^m
-\frac{2\pi i}{k_1}((\alpha -k_2\alpha_0)
_m s^m
-(\beta-k_2\beta_0)^m t_m)-\frac{i}{2k}J\right]
\ED
\BE \label{braid}
\cdot\Psi_{p,r+t}(0)
=\sum_{r^\prime}[B_{s,t}]_{r,r^\prime}\Psi_{p,r^\prime}(0)
\EE
with
$\sum_iq_i\int_{z_0}^{z_i(0)}\omega^l=\alpha_0^l+\Omega^{lm}\beta_{0m}$
and
where $J=\sum_i q_i^2(f_i(t)-f_i(0))$ is a self-linking term.

The self-linking contribution in (\ref{braid}) plays a very important role.
To construct a framing of the curve $z_i$, consider transporting a vector along
the
curve using a metric compatible connection on ${\cal M}$.  If this
operation were performed on an open manifold, on a curve $C$ enclosing
a region $U$, then the vector, after transport around the curve $C$,
would suffer a deflection by an angle given by a line integral of the
connection around this curve, which equals $\int_U R$, where $R$ is the
curvature of the connection on ${\cal M}$.  However, on a compact
Riemann surface this operation may equivalently be viewed as a
transport backwards along the curve $C$, encircling the complement
${\cal M} - U$, with the result $-\int_{{\cal M} - U} R$.  These
results are not the same, but differ by $\int_{\cal M} R = 4 \pi
(g-1) = 2 \pi \chi(g)$.  Our problem with this calculation is that the
Riemann surface can only be properly defined using a set of patches,
and viewed from different patches, parallel transport of a vector has
different meanings.  This ambiguity complicates further the framing
ambiguity already mentioned for which, we recall, a special choice,
a constant framing implying $J=0$,
made our theory into a topological field theory.  However, this choice
can be defined only on a single patch -- thus, for a consistent
topological field theory, we must ensure that the additional phase
ambiguity which arises in the wave function is rendered harmless, that is,
\BE \label{econs}
e^{\frac{2\pi}{k}i(g-1)q_i^2}=1
\EE
These are fundamental constraints, for $g\ne 1$, on the set of possible charges
$q_i$ depending on $k_2$ and $g$.

The matrices (\ref{braid}) satisfy the cocycle relation
\BE
B_{s_1,t_1}B_{s_2,t_2}=e^{-\frac{2\pi i}{k}(s_1^m{t_m}_2-s^m_2{t_m}_1)}
B_{s_2,t_2}B_{s_1,t_1}
\EE
(to be contrasted with the large gauge transformations cocycle
(\ref{clka})).  Together with the permutation phase $\sigma$, these
matrices are the result of the action of elements of the (permuted)
braid group on the particles which form the external sources in our
theory, and we might reasonably expect them to be a representation of
the braid group.  In fact, let the integer vectors $\hat s^l,\hat t_m$
denote vectors that are 0 in all entries except for the $l$th and
$m$th, respectively, and 1 at the remaining position.  Then with the
identifications $\alpha_l = B_{\hat s^l,0},\
\beta_m = B_{0,\hat t_m}$, it is easy to check that we recover all of
the conditions that appear in the work of Imbo and March-Russell
(\cite{4}), that is their equation (2.6).
We do not quite find their constraint (2.5), which restricts the value of
the phase
$\sigma$ as $\sigma^{2(n+g-1)} = 1$.
{}From (\ref{econs}), the above constraint is reduced to $\sigma
^{2n}=1$. Since they are considering a set of identical particles, we would
have to work in the sector of total charge $Q=nq$. Actually, this constraint
is a global condition on the representation of the braid group. Here we have
to suppose that there is an additional (or a group of) particle of charge $-Q$
present. The work in this non-zero
charge sector is under progress now and will be published in a later paper.
We expect to recover this additional constraint.

We have quantized Abelian Chern-Simons Theory coupled to arbitrary
external sources on an arbitrary Riemann surface, and solved the
theory.  We find that the presence of nontrivial spatial topology
introduces extra dimensionality to the Hilbert space separately for
the large gauge transformations and the braid group.
We find the constraints
(\ref{econs}), relating the
charges, $k$, and $g$ such that we recover a consistent topological
field theory representing the braid group on ${\cal M}$.

\end{document}